# Discovering electron transfer driven changes in chemical bonding in lead chalcogenides (PbX, where X = Te, Se, S, O)

S. Maier[1], S. Steinberg[2], Y. Cheng[1], Carl-Friedrich Schön[1], M. Schumacher[3], R. Mazzarello[3,4], P. Golub[5], R. Nelson[2], O. Cojocaru-Mirédin[1], J.-Y. Raty[6,7], M. Wuttig[1,4,8]*

[1] Institute of Physics IA, RWTH Aachen University, 52074 Aachen, Germany
[2] Institute of Inorganic Chemistry, RWTH Aachen University, 52056 Aachen, Germany
[3] Institute for Theoretical Solid State Physics, RWTH Aachen University, 52056 Aachen, Germany
[4] Jülich-Aachen Research Alliance (JARA-HPC), RWTH Aachen University, 52056 Aachen, Germany
[5] J. Heyrovsky Institute of Physical Chemistry, Department of Theoretical Chemistry, Dolejškova 2155/3, 182 23 Prague 8, Czech Republic
[6] CESAM and Physics of Solids, Interfaces and Nanostructures, B5, Université de Liège, B4000 Sart-Tilman, Belgium
[7] UGA, CEA-LETI, MINATEC campus, 17 rue des Martyrs, F 38054 Grenoble Cedex 9, France
[8] JARA-FIT Institute Green-IT, RWTH Aachen University and Forschungszentrum Jülich, 52056 Aachen, Germany

**Abstract:** Understanding the nature of chemical bonding in solids is crucial to comprehend the physical and chemical properties of a given compound. To explore changes in chemical bonding in lead chalcogenides (PbX, where X = Te, Se, S, O), a combination of *property-*, *bond breaking-* and *quantum-mechanical* bonding *descriptors* have been applied. The outcome of our explorations reveals an electron transfer driven transition from metavalent bonding in PbX (X = Te, Se, S) to iono-covalent bonding in β-PbO. Metavalent bonding is characterized by adjacent atoms being held together by sharing about a single electron (ES ≈ 1) and small electron transfer (ET). The transition from metavalent to iono-covalent bonding manifests itself in clear changes in these *quantum-mechanical descriptors* (ES and ET), as well as in *property-based descriptors* (i.e. Born effective charge ($Z^*$), dielectric function $\varepsilon(\omega)$, effective coordination number (ECON) and mode-specific Grüneisen parameter ($\gamma_{TO}$)), and in *bond breaking descriptors* (PME). Metavalent bonding collapses, if significant charge localization occurs at the ion cores (ET) and/or in the interatomic region (ES). Predominantly changing the degree of electron transfer opens possibilities to tailor materials properties such as the chemical bond ($Z^*$) and electronic ($\varepsilon_\infty$) polarizability, optical band gap and optical interband transitions characterized by $\varepsilon_2(\omega)$. Hence, the insights gained from this study highlight the technological relevance of the concept of metavalent bonding and its potential for materials design.

**List of keywords:** Metavalent bonding, thermoelectrics, phase change materials, chalcogenides, atom probe tomography

# Introduction

Understanding chemical bonding is of significant interest since it allows us to comprehend and tailor certain material properties [1], which could be utilized e.g. to optimize phase change materials [2] or thermoelectrics [3]. The first steps to understand the nature of the chemical bond were already taken almost a century ago by Linus Pauling [4] and others [5]. In the meantime, enormous developments have taken place in both, quantum-mechanical and experimental techniques [6], which help us to explore chemical bonding with unprecedented detail. Recently, these advances have also led to the concept of metavalent bonding (MVB), describing a bonding mechanism in between electron delocalization (*i.e.* metallic bonding) and electron localization at the ion cores (*i.e.* ionic bonding) as well as within the interatomic region (*i.e.* covalent bonding).[7]

Metavalent bonding has been categorized by combining both, *quantum-mechanical* and *experimentally accessible bonding descriptors*. [7] The *quantum-mechanical descriptors* are based on the electrons, which are shared and transferred between two atoms, i.e. they are related to the degree of electron (de-)localization in a given system. The *experimentally accessible descriptors* are the following measurable quantities: **(I)** the chemical bond polarizability measured by the Born effective charge ($Z^*$), **(II)** the optical dielectric properties as described by the dielectric function ($\varepsilon(\omega)$), including two of its characteristics, **(IIa)** the optical dielectric constant ($\varepsilon_\infty$), which is linked to **(IIb)** the optical band gap ($E_g$), **(III)** the effective coordination number (ECON), **(IV)** the electrical conductivity ($\sigma$), **(V)** a transverse optical mode-specific Grüneisen parameter ($\gamma_{TO}$) and **(VI)** the probability of multiple events (PME). **(I)**-**(V)** are *property-based*, while **(VI)** has been defined as a *bond breaking descriptor* [7c], which is why **(VI)** is discussed separately. Materials, which comprise the metavalent type of bonding, have been designated as 'Incipient Metals'. They show significant differences in these descriptors relative to conventional covalent, ionic, and metallic systems.[7a, 7b]. The logic consequence of these findings was to define another fundamental bonding type. Categorizing the diverse types of bonding by means of the aforementioned descriptors has provided fundamental insights on the nature of metavalent bonding and on how it differs from covalent, ionic and metallic bonding. In particular, the border between metavalent and covalent bonding has been studied recently. [8,9] It was shown that several bonding descriptors, including $Z^*$ and $\varepsilon_\infty$ as well as the bond breaking (i.e. the PME) showed a discontinuous change upon the transition from metavalent to covalent bonding. These findings confirm that metavalent bonding is a distinct fundamental type of bonding which differs substantially from covalent bonding. This immediately raises the question, if it is possible to identify and study a transition and its nature between metavalent and ionic bonding.

In order to answer this question, we studied the aforementioned *bonding descriptors* for a series of materials, i.e. PbX (X = Te, Se, S, O), which we anticipated to be located at the boundary between metavalent and iono-covalent bonding. Since the tellurides of the post-transition-metals such as PbTe are often used in devices for thermoelectric energy conversion [10] it is of significant technological relevance to understand the nature of bonding of these materials even beyond the area of phase change materials, where metavalent bonding was discussed first. [7a] In fact, the concept of metavalent bonding can be considered relevant in all areas, where the degree of electron (de-)localization plays an important role.

Therefore, a combination of atom probe tomography, optical measurements and *quantum-mechanical* methods was used to explore the transition from metavalent to iono-covalent bonding with special emphasis on the role of electron transfer in PbX (X = Te, Se, S, O). Although the gross structural features of PbO differ from those of PbX (X = Te, Se, S), we will demonstrate that the aforementioned *bonding descriptors* allow distinguishing bonding types for materials with different crystal structures. In the following, we present the outcome of our explorations.

## Results and Discussions

### Crystal structures of PbX (X = Te, Se, S, O)

Lead monoxide exists in two different polymorphs – the α and β form. This study focuses mainly on β-PbO. Previous work [11] on the crystal structures of PbX (X = Te, Se, S, O) revealed that β-PbO crystallizes in the massicot structure, where each lead atom is coordinated by four oxygen atoms. The same coordination number, but a different atomic arrangement is found in α-PbO (litharge). On the contrary, the higher lead chalcogenides (PbX, where X = Te, Se, S) crystallize in the rocksalt-type structure, where the atoms are octahedrally coordinated. The different crystal structures are illustrated in **Fig 1**. It is by itself remarkable that PbTe, PbSe and PbS crystallize in the rocksalt-type structure, which is frequently encountered for ionic materials. The physical properties of these materials however, are far from what one would expect for an ionic material (e.g. small band gaps, metallic luster, relatively high carrier concentrations etc.). These materials are also not fulfilling the 8-N rule. Therefore, a simple covalent description of chemical bonding is not sufficient either. With this work, we shed light on this peculiar bonding situation in the context of metavalent bonding with emphasis on the dominant role of electron transfer.

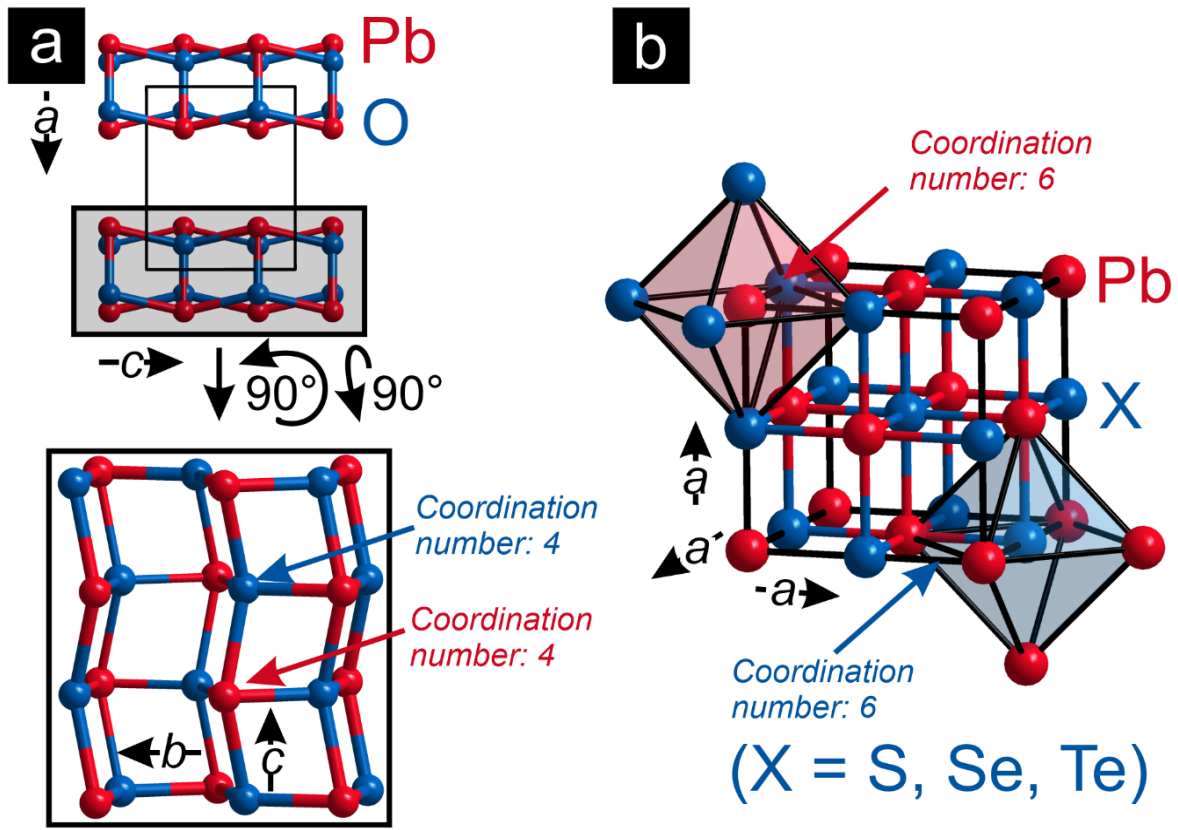

**Fig. 1:** Representations of the crystal structures of **(a)** the massicot-type β-PbO and **(b)** the rock salt-type PbX (X = S, Se or Te): the diverse types of coordination polyhedrons are shown in the insets. Interestingly, in massicot-type PbO the four oxygen ions surrounding a Pb cation are all located on one side. The hypothetical rock salt-type PbO phase is equivalent to the structural model shown in (b).

### *Property-based descriptors for chemical bonding*

In the following, the *property-based bonding descriptors (cf.* **Fig. 2***),* which were used to explore the transition between different types of chemical bonding, are introduced and discussed. The optical descriptors are the dielectric constant ($\varepsilon_\infty$), linked to the electron polarizability, the Born effective charge ($Z^*$), which characterizes the chemical bond polarizability, the optical band gap ($E_g$) as well as the mode-specific Grüneisen parameter ($\gamma_{TO}$) of the optical phonons, which is an indicator for the softness of the chemical bonds. The atomic arrangement (i.e. the effective coordination number) in a crystal structure can be used to differentiate between metallic bonding (characterized by a large number of nearest neighbors), covalent bonding (8-N rule) and metavalent bonding, where the effective coordination number differs significantly from the 8-N rule. The Born effective charge ($Z^*$) is defined according to (1). For anisotropic materials, $Z^*$ is a tensor. To simplify the problem, we made the assumption that $Z^*$ is isotropic, which results in a scalar. Since β-PbO is not cubic, we also provide the whole dielectric and Born effective charge tensors in the supplement.

$$Z^* = \left(\frac{dP(u)}{du}\right) \quad (1)$$

Equation (1) describes the change in polarization upon an atomic displacement, related to a change in electron wave functions, which induces an electric dipole moment (P) in the direction of the displacement (u). [12] This makes Z* a good bonding descriptor, describing the polarizability of chemical bonds.

**Fig. 2a** shows an anomalously high chemical bond polarizability Z* for PbTe, PbSe and PbS – one characteristic property of incipient metals. In order to ensure comparability between different compounds, Z* was normalized by the formal oxidation state of the constituents ($Z_n^*$). $Z_n^*$ decreases when going from PbTe to PbS, i.e. the charges become more localized and the bonds less polarizable. **Fig. 2b-c** show a concomitant gradual decrease in $\varepsilon_\infty$ and an increase in $E_g$, which are both related, as expressed for example in the Moss rule. [13] All band gaps were obtained from DFT computations including spin-orbit coupling (SOC) (*cf.* **Fig.S1** in the supplement). The strongest changes are observed between PbS and β-PbO. While the properties of β-PbO are perfectly in line with ordinary iono-covalent bonding (e.g. $Z_n^* \approx 1$) the large values of $Z_n^*$ and the mode-specific Grüneisen parameter for transverse optical modes ($\gamma_{TO}$) together with the high value of $\varepsilon_\infty$ and the electrical conductivity found for the higher chalcogenides are neither characteristic for metallic, ionic nor for covalent bonding. Instead, all properties observed can be explained by the concept of metavalent bonding. [7a,14] Hence, the transition between metavalent and iono- covalent bonding proceeds between PbS and β-PbO. Unfortunately, studying the transition between PbS and β-PbO in more detail by mixing PbS and PbO is not possible experimentally due to decomposition and generation of toxic $SO_x$. Nevertheless, further experiments and DFT calculations can be performed to unravel the nature of the transition of properties and bonding going from PbTe to PbO. To consider the role of the different chemical environments, the effective coordination numbers (ECONs) were inspected. [15] These are good structural descriptors for chalcogenides, where frequently distortions (e.g. Peierls distortion) away from a high-symmetry atomic arrangement are observed. Such distortions are a result of a competing electron (de-) localization characterized by ES and are only relevant, when ES instead of ET (as in this case) dominates the transition. Yet, for PbX (X = Te, Se, S) no such distortion is found. Between PbS and β-PbO the chemical environment changes from ECON = 6 to ECON = 4 (*cf.* **Fig. 2d**). Hence, the observed change in ECON between PbS and PbO might be related to changes in charge transfer.

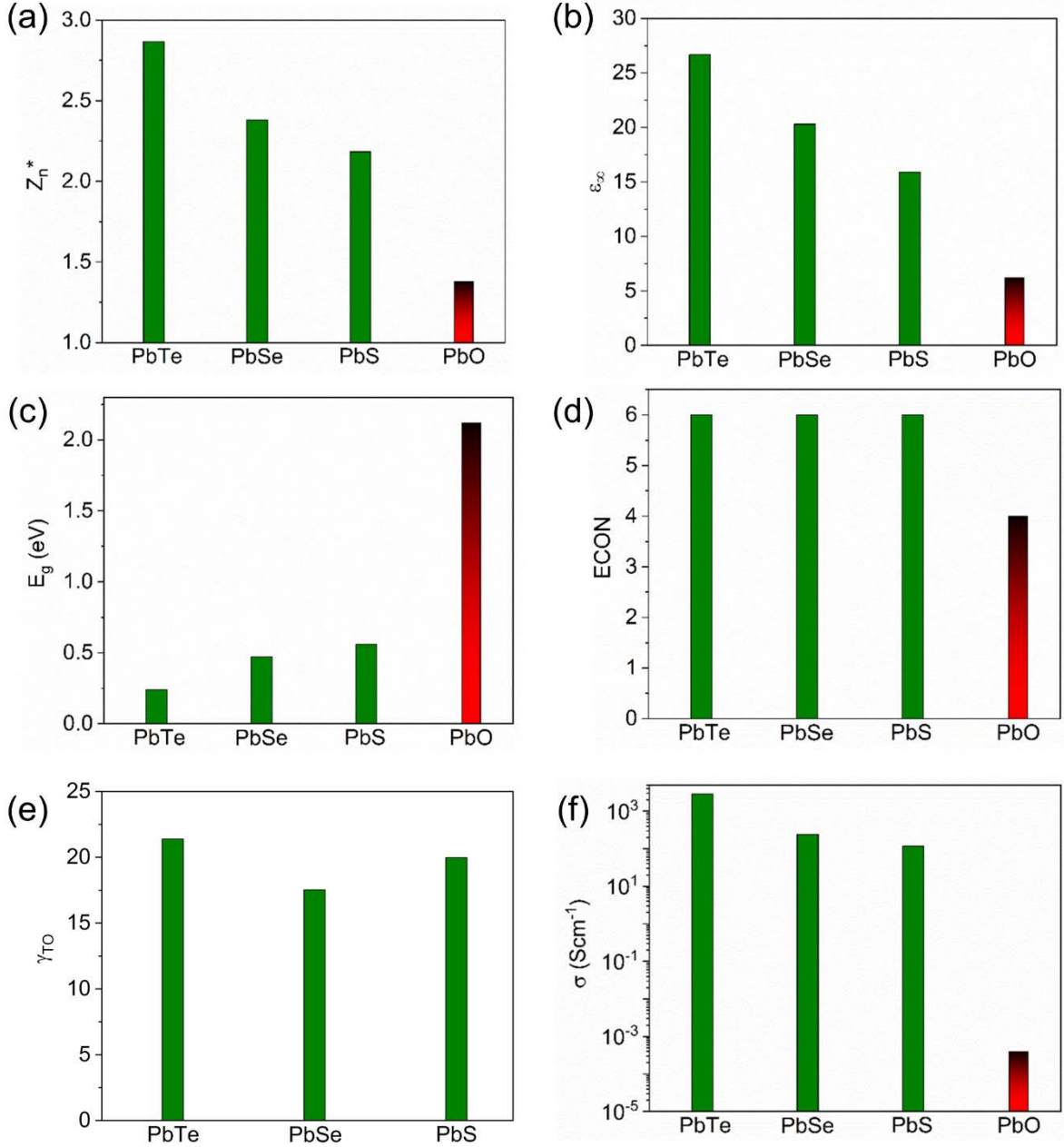

**Fig. 2: (a)** Normalized Born effective charge $Z_n^*$, **(b)** optical dielectric constant, **(c)** optical band gap, **(d)** effective coordination number (ECON) of Pb, **(e)** mode-specific Grüneisen parameter and **(f)** electrical conductivity of PbX (X = Te, Se, S, O). Clearly, the property portfolio of β-PbO differs substantially from the property combination characterizing the three higher lead chalcogenides. We attribute this to a change in bonding mechanism going from metavalent to iono-covalent bonding, as will be further substantiated below.

The anharmonicity (or softness) of the bonds is determined by one mode-specific Grüneisen parameter ($\gamma_{TO}$), which describes how the frequency of the transverse optical modes changes with the cell volume. **[16]**

**Fig. 2e** shows that the bonds are very soft for the higher chalcogenides (PbS to PbTe). For β-PbO, the unit cell is very large, so that a large number of optical modes exist. In this case, it is simpler to characterize the bond stiffness/softness by the Debye temperature as also demonstrated for the sesqui-chalcogenides. [8] The degree of electron (de-)localization also has an influence on the electrical conductivity. It can be seen from **Fig. 2f** that the conductivity in β-PbO is several orders of magnitude lower compared to the higher chalcogenides, which is indicative for a change in the degree of electron (de-)localization, which will be further discussed later on. The intrinsic electrical conductivity already decreases when going from PbTe to PbS (*cf.* **Fig. S2**), i.e. charges apparently become more localized, which is in line with an increase in electron transfer. All observed changes in physical properties can be explained by a difference in chemical bonding between β-PbO and the other three lead chalcogenides. Yet, this change of properties is also accompanied by a change in atomic arrangement. Hence, it is desirable to find further arguments for a change in bonding independent of the nature of the crystal structure. To reach this goal, we now shift our focus towards the probability of multiple events (PME), which is discussed in the next section.

*The probability of multiple events (PME) – A bond breaking descriptor for chemical bonding*

Laser-assisted atom probe tomography (APT) has been used recently to study the bond breaking behavior of various chalcogenides. [3a,7c, 8] Generating an electric field at the apex of a needle-shaped specimen by applying a DC voltage of 3 - 8 kV and in conjunction with a short laser pulse with an energy of 10 - 20 pJ enabled the observation of an unconventional bond breaking behavior for all metavalent materials studied so far. More specifically, an unusually high PME (probability of multiple events) has been systematically detected, while conventional materials show a low PME (values higher than 50% for MVB materials and lower than 30% for non-MVB materials). Hence, a change in bond breaking behavior can be detected by studying the field evaporation behavior with special focus on the PME. Such an analysis is employed and discussed here for the PbX (X = Te, Se, S, O) system. The results are summarized in **Fig. 3**. The following discussion focuses on two aspects: changes in the overall PME as well as the spatial distribution of the multiple events, i.e. the possible origin of the high PME. The first part allows conclusions concerning changes in bond breaking behavior, while the second one allows to exclude certain measurement artifacts such as DC and burst evaporation, which could also lead to a high PME. A detailed discussion concerning multiple events in APT analyses can be found in the supplement. It also includes a discussion about the impact of stoichiometry and evaporation behavior of the constituents on the PME of the compounds studied herein.

The PME of PbX (X = Te, Se, S, O) shows a drastic, step-like change between PbS and β-PbO. **Fig.3a** depicts a clear difference in how the bonds break, i.e. the PME suddenly drops from 64 % to 18 % (for more details, see discussion in the supplement) when going from PbS to β-PbO, which is also indicative for a change in chemical bonding and consistent with the observed property changes. These findings are supported by observed changes in the field evaporation behavior of the Pb ions, which are discussed in the supplement (*cf.* **Fig. S3**).

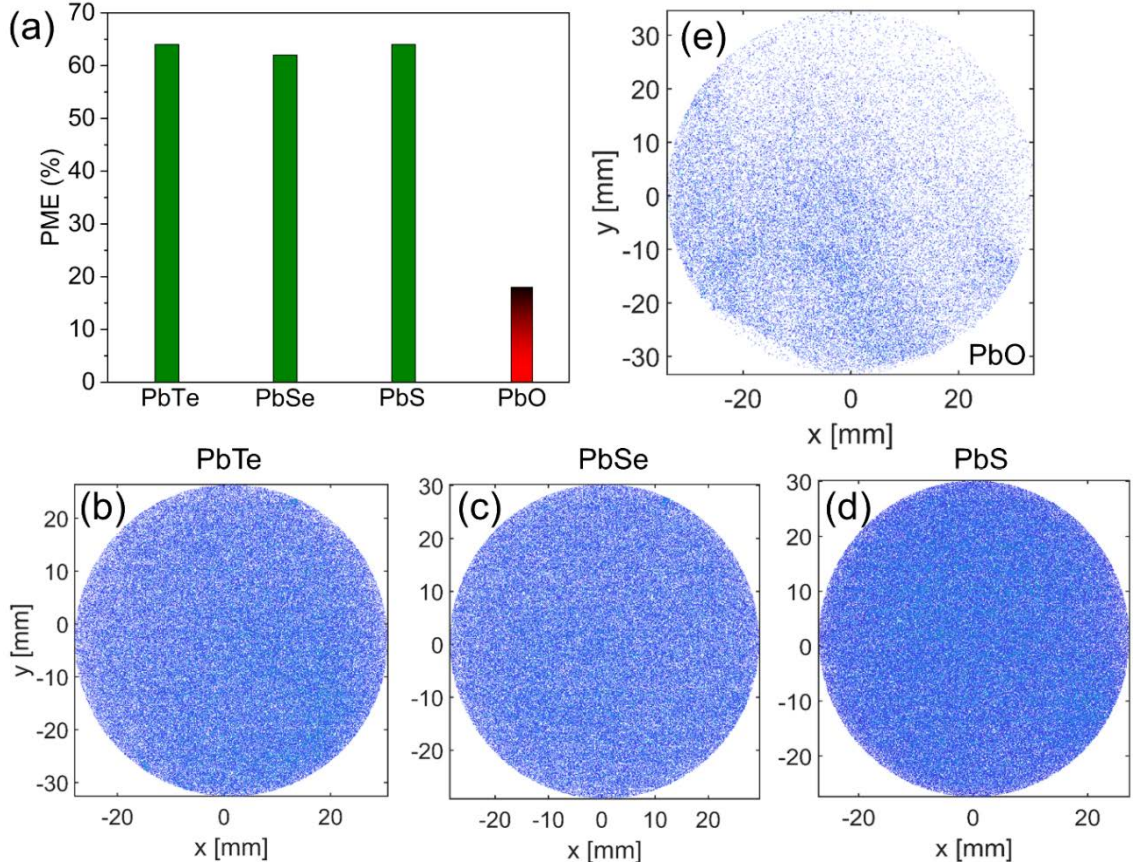

**Fig. 3: (a)** Probability of multiple events: 64 % for PbTe, 62 % for PbSe, 64 % for PbS and 18 % for β-PbO; Spatial distribution of multiple detector events of **(b)** PbTe, **(c)** PbSe, **(d)** PbS and **(e)** β-PbO on the atom probe detector. A much lower density and number of multiple events can be observed for β-PbO compared to the other PbX (X = Te, Se, S) compounds, which is also reflected in the much lower overall probability of multiple events of the datasets shown in **(a)**. Note that the size of the datasets in **(b)**-**(e)** was kept fixed to $5 \times 10^6$ ions. Yet, we display here only the multiple events clearly proving their superior number in PbTe**(b)**, PbSe **(c)**, and PbS when compared to β-PbO **(e)**. In the case of β-PbO, the spatial distribution of the multiple events is slightly inhomogeneous. The reason is a slightly inhomogeneous field evaporation, probably since β-PbO is very sensitive to the laser energy, i.e. if parts of the tip are illuminated more than others, inhomogeneous field evaporation can occur. Such an effect is well known for oxides such as ZnO, especially for higher laser pulse energies. **[17]**

The spatial distribution of the multiple events on the detector (*cf.* **Fig.3 b-e**) changes significantly between PbS and β-PbO in the sense that a much lower density of multiple events is detected for β-PbO compared to the higher chalcogenides, which all show a similar spatial distribution of multiple events. The multiple events observed for β-PbO result from ordinary correlated field evaporation – an effect inherent to every atom probe experiment. Please note that the size of the datasets is the same (i.e. $5\times10^6$ ions), which allows comparing the spatial distributions of multiple events of all four compounds. It is also noteworthy that the high density of multiple events in PbX (X = Te, Se, S) is homogeneous over the whole detection range of the detector, which is fundamentally different when comparing it to the elevated multiple events, e.g. observed for carbon in the body centered cubic ferrite (α-Fe). [18] There, the multiple events of carbon occur in segregated regions of low or high atomic density corresponding to poles and zone lines, i.e. they are not distributed homogeneously. There is also no sign of burst or DC evaporation in the analyzed datasets, which can be seen from the detection rates and correlation histograms shown in **Fig. S5** and **Fig. S6**. Hence, the observed differences in multiple events between β-PbO and PbX (X = Te, Se, S) can be considered intrinsic and are indicative for a difference in chemical bonding, consistent with the changes in physical properties (*cf.* **Fig. 2**).

<u>Electron-sharing and electron transfer in the context of metavalent bonding</u>

Creating materials property maps based on chemical quantities and maps to distinguish metallic, covalent and ionic bonding has a rather long history. Examples are the Van Arkel-Ketelaar triangle [19] or maps for phase-change materials based on hybridization and ionicity [20] as well as sp$^3$ mixing and ionicity. [21] Recently, a *quantum-mechanical* map for bonding in solids was introduced. [7b], which is based on two *quantum-mechanical* coordinates – the electrons shared (ES) and transferred (ET) between adjacent atoms. Remarkably, such a map is able to separate different bonding mechanisms as can be seen from **Fig.4**. In this study, we combine *quantum-mechanical*, *bond breaking* and *property-based bonding descriptors* such that different bonding mechanisms can be distinguished. The "backbone" of this approach is the electrons shared and transferred between adjacent atoms (*cf.* **Fig.4**) and it has already been used successfully to study the boundary between metavalent and covalent bonding. [8,9] In this work we focus on the boundary between metavalent and iono-covalent bonding. From **Fig. 4** it is already visible that covalent (red), metallic (blue) and ionic bonding (black) can be distinguished. In covalent materials electron-sharing is dominant (i.e. ES approaches 2, which corresponds to the classical Lewis electron pair), while in ionic compounds chemical bonding is dominated by electron transfer (ET approaches 1). In metallic systems the electrons are fully delocalized, i.e. charge localization through electron-sharing and electron transfer is minimized (i.e. ES and ET are small).

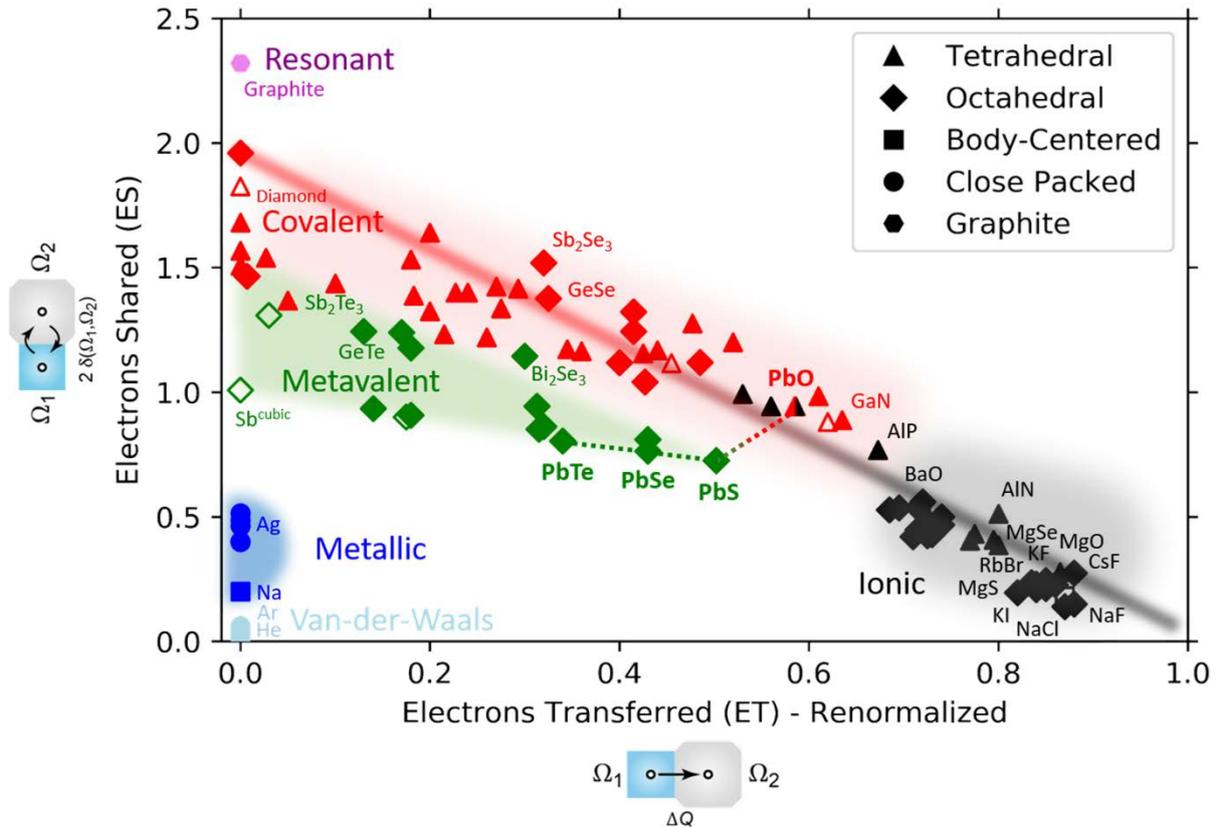

**Fig. 4:** A *quantum-mechanical* map for bonding in solids with its two coordinates: the number of electrons shared and transferred between adjacent atoms. The blue, red, black and green regions correspond to areas, where metallic, covalent, ionic and metavalent compounds are located. β-PbO falls right at the transition between covalent and ionic, while PbX (X = Te, Se, S) are all lying on a line within the metavalent region.

The metavalent region is located in between covalent, metallic and ionic bonding, where approximately one electron is shared between adjacent atoms and where electron transfer is moderate. We already discussed changes in property trends and bond breaking in the PbX (X = Te, Se, S, O) system. Now, we assign the four compounds to the different regions of the map by determining the number of electrons shared and transferred between adjacent atoms. It can be seen from **Fig. 4** that the higher chalcogenides are all lying on a line within the metavalent region, where (so far) all cubic, metavalent materials can be found. β-PbO, on the contrary, falls right at the transition point between covalent and ionic bonding, which is why we call it an ionocovalent compound throughout this manuscript. It is remarkable that the *quantum-mechanical* map for bonding in solids (*cf.* **Fig. 4**) predicts a transition from metavalent to iono-covalent between PbS and β-PbO, where we also observe the previously discussed changes in the *property-based* and *bond breaking descriptors*.

Which consequences do these findings have for our understanding of metavalent bonding in general? We now know that metavalent bonding has at least two well-defined borders to the covalent and iono-covalent regime of the map. This work demonstrates that metavalent bonding can be weakened not only by increased electron-sharing (as shown previously [**8,9**]), but also through an increase in electron transfer, where the former can result in structural distortions (e.g. Peierls distortion). Hence, changes in chemical bonding cannot only be induced by increasing electron-sharing as it was recently achieved for some $V_2VI_3$ compounds [**8,9**], but also by electron transfer. Our work therefore demonstrates that there are two fundamental strategies to tune chemical bonding quantitatively: changing the degree of electron-sharing and electron transfer. This casts doubts on theories only focusing on one parameter when discussing systematic changes in chemical bonding. In the next section we will demonstrate that these fundamental parameters (electrons shared (ES) and electrons transferred (ET)) can also be used to systematically tailor relevant material properties.

<u>Electron-sharing and electron transfer in the context of materials design</u>

The holy grail in most areas of materials science is the design and tailoring of functional materials and their respective properties. In this section we will demonstrate that using ES and ET can be an effective approach towards materials design. This can be seen from **Fig. 5** and **Fig. S8**, which show that by controlling ES and ET one can systematically tailor physical properties such as $E_g$, $\varepsilon_\infty$, $Z^*$ and $\gamma_{TO}$. In the case of the PbX (X = Te, Se, S and O) system, $Z^*$ and $\varepsilon_\infty$ decrease, while $E_g$ increases systematically upon increasing electron transfer. Other quantities such as ECoN, $\gamma_{TO}$ and $\sigma$ change significantly in the vicinity of the transition from metavalent PbS to iono-covalent β-PbO (also see **Fig. 2**). Importantly, the trends observed in figure 5 appear to be generic for all incipient metals. Increasing ES reduces ECoN, $\varepsilon_\infty$ and $\gamma_{TO}$, while ET also opens the band gap and has an effect on $\varepsilon_\infty$. Hence, it is possible to systematically tailor optical and other relevant material properties via electron transfer. This will also be demonstrated in the following using the example of the imaginary part of the dielectric function ($\varepsilon_2(\omega)$) in the region of the electronic interband transitions. For PbTe, PbSe and PbS, $\varepsilon_2(\omega)$ was extracted from experimental elipsometry data (*cf.* solid black lines in **Fig. 6 a-c**). Details concerning thin film preparation and analyses (XRR, XRD and elipsometry) can be found in the experimental section and in the supplement. We observed that systematically increasing the degree of electron transfer has three effects: The maximum of $\varepsilon_2(\omega)$, $\varepsilon_2^{max}$, shifts towards higher energies and decreases in magnitude, while $\varepsilon_2(\omega)$ broadens. We confirmed these observations using DFT calculations (details can be found in the experimental section).

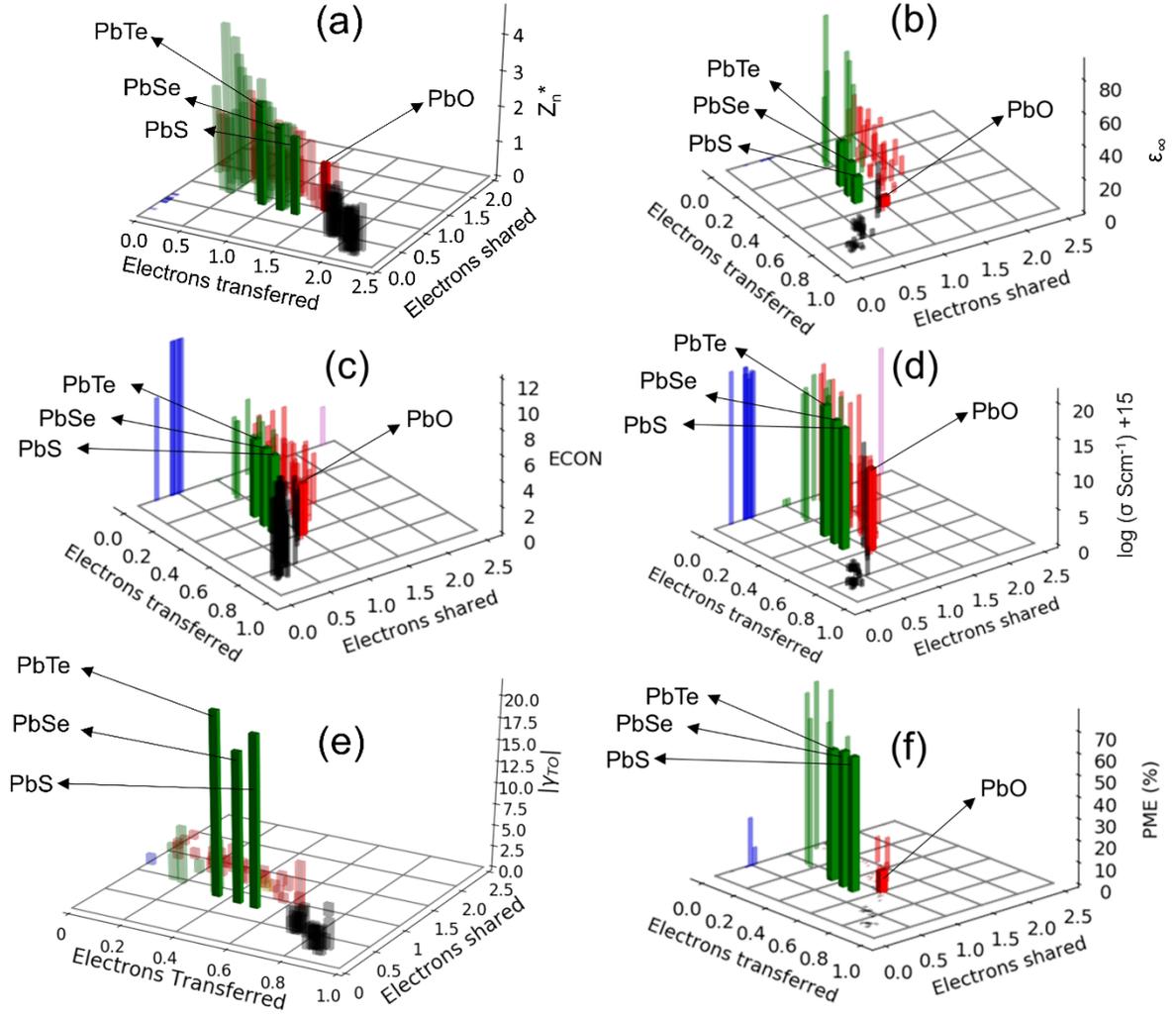

**Fig. 5:** (**a**) Born effective charge (normalized by oxidation state), (**b**) optical dielectric constant, (**c**) effective coordination number (ECoN), (**d**) electrical conductivity (logarithmic scale), (**e**) mode-specific Grüneisen parameter and (**f**) probability of multiple events (PME) of PbX (X = Te, Se, S, O) and other solid state materials, correlated with the electrons shared and transferred between adjacent atoms. This figure clearly demonstrates that by controlling ES and ET one can systematically tailor physical properties, such as $E_g$, $\varepsilon_\infty$, $Z^*$, $\sigma$ and $\gamma_{TO}$.

The theoretical results reproduce the trends observed experimentally. Furthermore, as shown in **Fig. 6** and **Fig. S12**, they allow to identify the contributions of the different states as indicated by the different color code, enabling further insight into the observed changes in $\varepsilon_2(\omega)$. The simulations clearly show that the electronic interband transitions between 0 eV and 10 eV stem from p-p transitions (green). Other contributions become relevant at higher energies, which can be seen in **Fig. S12** in the supplement.

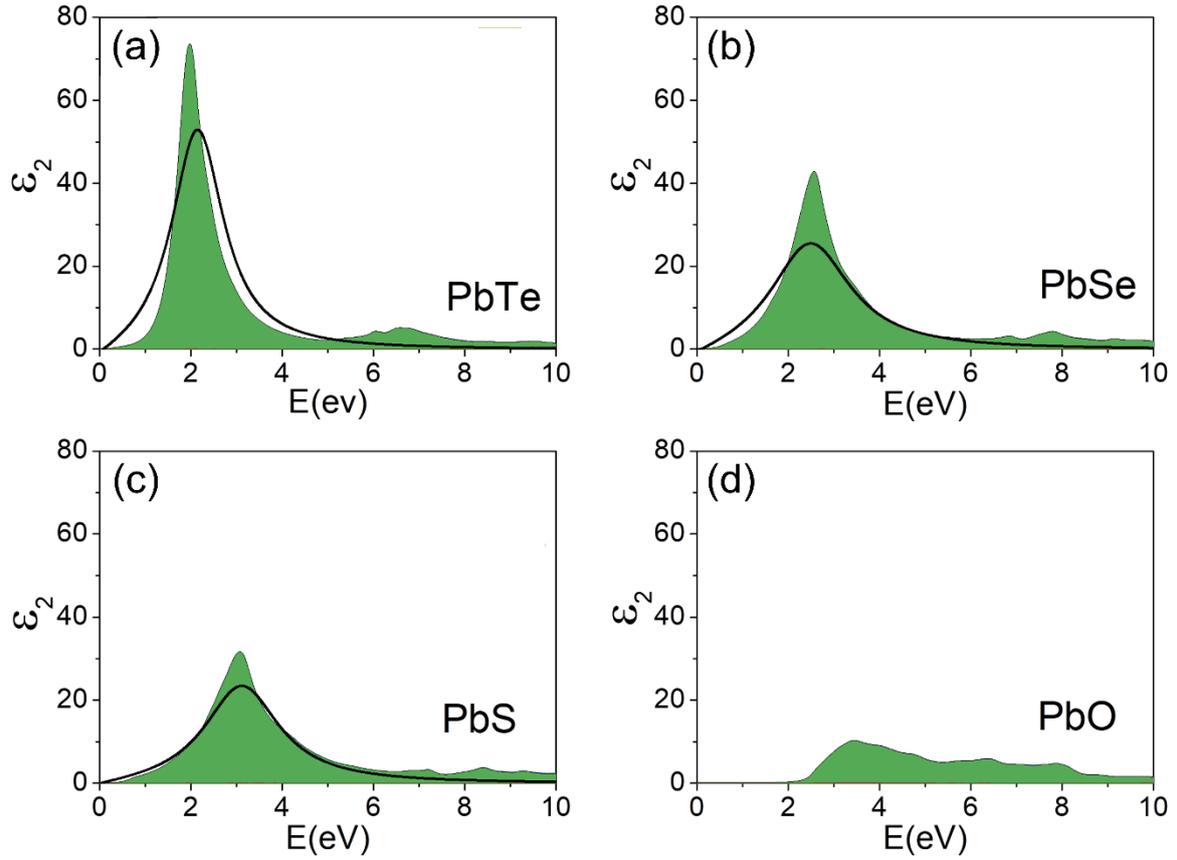

**Fig. 6:** Imaginary part of the dielectric function ($\varepsilon_2(\omega)$) in the region of the electronic interband transitions. Experimental data are depicted as solid black lines for PbTe, PbSe and PbS. It can be observed that the maximum of $\varepsilon_2(\omega)$, $\varepsilon_2^{max}$, shifts towards higher energies and decreases in magnitude, while $\varepsilon_2(\omega)$ broadens. These observations were confirmed using DFT calculations. The only contribution to the electronic interband transitions between 0 eV and 10 eV arises from p-p transitions (green).

The shift of $\varepsilon_2^{max}$ towards higher energies can be attributed to the opening of the optical band gap (*cf.* **Fig. 2**). In order to understand the decrease in $\varepsilon_2^{max}$ as well as the broadening of $\varepsilon_2(\omega)$ and how these phenomena are connected to changes in chemical bonding one has to decompose $\varepsilon_2(\omega)$ into its contributions. These are the transition probability ($W_{i,f}$) from the initial to the final state as well as their joint density of states (JDOS). The overall transition probability for all allowed interband transitions ($W_{i,f}$) depends on the corresponding matrix element (ME) of the interband transitions and the JDOS, which can both affect the magnitude of $\varepsilon_2^{max}$ and shape of $\varepsilon_2(\omega)$. When going from PbTe to PbS electron transfer increases without a structural transition. In this case, the JDOS do not change significantly, which can be seen from **Fig. 7**. The logic conclusion, which can be drawn from these findings is that the changes in magnitude of $\varepsilon_2^{max}$ and shape of $\varepsilon_2(\omega)$ observed for PbTe, PbSe and PbS are mainly caused by differences in the matrix element (*cf.* **Fig. 7**).

This raises one important question. How can the matrix element be connected to chemical bonding and electron transfer? The magnitude of the matrix element depends on orbital parity and the degree of orbital overlap. In the case of the PbX (X = Te, Se, S, O) system, symmetry allowed p-p transitions dominate (*cf.* **Fig. 6**). A decreasing p-orbital overlap will result in a reduced matrix element of the corresponding transitions, which in turn reduces the magnitude of $\varepsilon_2^{max}$ and influences the shape of $\varepsilon_2(\omega)$ as observed (*cf.* **Fig.7**). The three isostructural PbX (X = Te, Se, S) compounds, show a significantly larger p-p orbital overlap than β-PbO, which can explain the rather low magnitude of $\varepsilon_2^{max}$ and broad shape of $\varepsilon_2(\omega)$ (*cf.* **Fig. 6**). What is rather remarkable is the apparent decrease in $\varepsilon_2^{max}$ and broadening of $\varepsilon_2(\omega)$ upon systematically increasing electron transfer within the isostructural PbX (X = Te, Se, S) series. These observations suggest that orbital overlap and hence optical matrix element can be systematically tuned by predominantly changing ET.

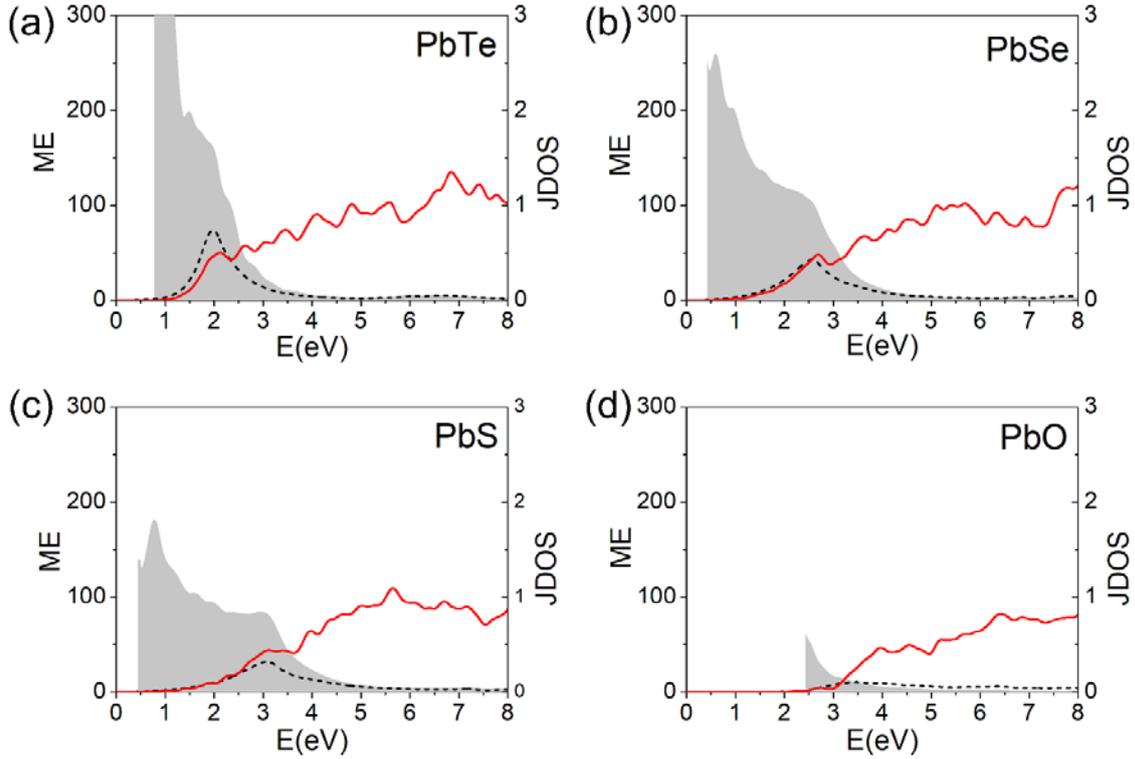

**Fig. 7:** Matrix element (ME) and joint density of states contributing to the total imaginary part of the dielectric function $\varepsilon_2(\omega)$ in the energetic region of the electronic interband transitions shown in **Fig. 6**. Left axis: matrix element (ME, grey area) and total $\varepsilon_2(\omega)$ (dashed line); right axis: joint density of states (red, DOS normalized per atom). When going from PbTe to PbO, the decrease in $\varepsilon_2^{max}$ and broadening of $\varepsilon_2(\omega)$ can be attributed to changes in the matrix element ME causing a change in the electronic interband transition rate. The reduced transition rate can be explained by a reduced orbital overlap between initial and final state. This reduction is related here to the increasing charge transfer, as described by ET. On the contrary, in GeTe the optical contrast is reduced by a reduction of the relative orientation of the p-orbitals of adjacent atoms as described by [22].

## Metavalent bonding in the context of technological applications

This work and previous studies [3,7,8,9] have revealed that changing the degree of electron-sharing and electron transfer has significant implications for the physical properties of a fairly large number of inorganic materials. This study in particular demonstrates that it is possible to systematically tailor the optical (and other relevant) properties of solids by predominantly changing the degree of electron transfer (*cf.* **Fig. 5-7**). This is especially evident from the property trends within the isostructural PbX (X = Te, Se, S) series, presented here. Increasing electron transfer reduces the chemical bond polarizability ($Z^*$) and electronic polarizability ($\varepsilon_\infty$), as well as the optical interband transitions characterized by $\varepsilon_2(\omega)$ and increases the optical band gap (*cf.* **Fig.6-7**). These findings can be summarized in an interactive materials map. [23]

Even though these studies [3,7,8,9] are fundamental in nature, it is important to put their findings in a technological context. As a start we restrict this discussion to a selected number of metavalent materials, which are already technologically relevant. These are for example PbTe, $Bi_2Te_3$ and $Ge_2Sb_2Te_5$ as well as other GST-based phase-change materials. [7c] Since this study focuses on the PbX (X = Te, Se, S, O) system, we discuss our findings in the context of thermoelectric applications. However, it is important to keep in mind that the recipe of systematically tailoring the optical properties developed here, also provides a new avenue for designing phase change materials. PbTe and PbSe are known to be excellent thermoelectric materials and many studies focused on different aspects playing a role in their good performance. [24] This study highlights the importance of the unique degree of electron-sharing and transfer characteristic for metavalent materials, which is also important for the thermoelectric performance since it determines the magnitude of the Seebeck coefficient (S = 0 for metals, i.e. the electrons are too delocalized for a high performance) and the electrical conductivity (insulators, where the electrons are too localized, possess a large Seebeck coefficient, but have too low electrical conductivities to compete with high-performance thermoelectrics). [3a, 25]

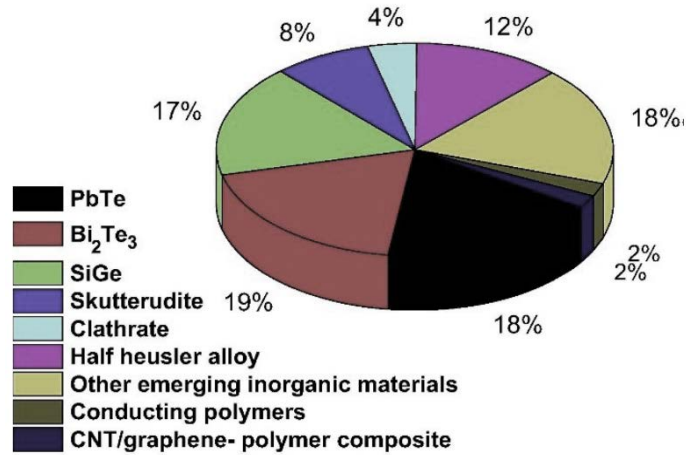

**Fig. 8** Thermoelectric materials used in research and development. Figure adapted from Ref. [26].

PbTe and $Bi_2Te_3$ (also metavalent) [8] are the two leading thermoelectric materials in both research and development (*cf.* **Fig. 8**). Hence, systematically developing new inorganic PCMs and thermoelectric materials using the strategy proposed here, can be of significant technological interest. Further research is therefore recommended to focus on developing new PCMs and thermoelectric materials using the concept of metavalent bonding in order to assess its technological relevance and its potential for materials design in more detail and a broader scope.

## Conclusion

The systematic explorations of the PbX (X = Te, Se, S, O) system presented herein, clearly reveal an electron transfer driven change in chemical bonding, which occurs between PbS and β-PbO. The nature of bonding in the oxide is iono-covalent, while PbX (X = Te, Se, S) are identified as metavalently bonded 'incipient metals'. The transition from metavalent to iono-covalent bonding is achieved by predominantly changing the degree of electron transfer. This is fundamentally different for the transition metavalent-covalent, which is dominated by changes in the degree of electron-sharing. This work demonstrates that predominantly changing the degree of electron transfer opens possibilities to tailor materials properties such as the chemical bond ($Z^*$) and electronic ($\varepsilon_\infty$) polarizability, optical band gap and optical interband transitions characterized by $\varepsilon_2(\omega)$. The insights, gained from this study, therefore provide a recipe for future assessments of the technological relevance of the concept of metavalent bonding and its potential for materials design.

**Experimental Section**

*Sample preparation:* PbTe and PbSe bulk samples were synthesized from reactions of the pure elements. [27] In order to verify the purity of these samples, powder X-ray diffraction patterns were recorded at room temperature using a STADI P diffractometer equipped with an area detector and a Cu-Kα source. **Fig. S7** in the supplementary information shows that both samples are phase pure within the X-ray detection limit. PbS lumps and PbO powder were purchased from Sigma Aldrich® (purity: 99,995% for PbS and 99.999% for PbO) and used without further purification. PbTe, PbSe and PbS thin films were deposited on single side polished Si <100> substrates for the corresponding elipsometry measurements in order to extract $\varepsilon_2(\omega)$ experimentally (*cf.* **Fig.6**). PbTe and PbSe films were deposited using DC sputtering (base pressure 2 x $10^{-6}$ mbar, P = 20 W, Ar-flow: 20 sccm), while PbS was sputtered using a RF generator (base pressure 2 x $10^{-6}$ mbar, 60 W, 20 sccm) due to its higher resistance. In all cases commercial, stoichiometric targets with a purity of 99.99 % were used. The film structure was verified by X-ray diffraction (*cf.* **Fig. S9** in the supplement), while the film densities were determined using X-ray reflectivity measurements (*cf.* **Fig. S10** in the supplement).

*Atom probe tomography (APT):* Needle-shaped samples with a radius of 50-100 nm were prepared using a standard two-step process, i.e. sample lift-out and subsequent annular milling. Lift-outs for β-PbO were done from the corresponding powder particles. A FEI® Helios Nanolab 650 dual-beam and commercially available flat top Si microtips were used for all sample preparations. Subsequently, the APT samples were immediately transferred into an ultra-high vacuum chamber (p ≈ $10^{-11}$ mbar) to avoid surface oxidation. All APT measurements were performed employing a LEAP 4000 X Si local electron atom probe (Cameca® Instruments) in laser pulse mode ($\lambda_{laser}$ = 355 nm) at a base temperature of 55 K. A laser pulse energy of 20 pJ, a pulse frequency of 200 kHz and a detection rate of 1 % were used for all measurements. The resulting data sets were analyzed using the IVAS 3.6.12 software package. A detailed analysis of the multiple events was performed using the EPOSA software package, which was developed in-house.

*Elipsometry:* Elipsometry spectra ranging from 0.7 to 5.2 eV and for angles of incidence of 65°, 70° and 75° were recorded using a Woolam M-2000UI elipsometer equipped with deuterium and halogen lamps. Experimentally obtained elipsometry data were fitted using the CODE software (*cf.* **Fig. S11**). The dielectric function $\varepsilon(\omega)$ was modeled using the following contributions: (1) background, which is a constant accounting for the polarizability of the higher energy range, (2) a Drude contribution and (3) a Tauc–Lorentz model. The corresponding fits of the experimental elipsometry data in the energy range of interest to compare experimental and theoretical $\varepsilon_2(\omega)$ (*cf.* **Fig. 7**) can be found in the supplement.

*Computational Details:* Full structural optimizations (lattice parameters and atomic positions) and electronic structure computations were performed using Density Functional Theory in combination with projector augmented wave [28] (PAW) potentials as implemented in the *Vienna ab initio simulation package* (VASP)[29] and *ABINIT* [30] code, respectively. Correlation and exchange were described by the generalized gradient approximation of Perdew, Burke, and Ernzerhof (GGA–PBE)[31], while the density-functional-theory-based computations were performed considering spin-orbit-coupling. Sets of 5 × 5 × 5 up to 20 × 20 × 20 *k*-points were employed to sample the first Brillouin zones for reciprocal space integrations. The energy cut-off was 500 eV and all calculations were considered converged, as the difference between two iterative steps of the electronic (and ionic) relaxations fell at most below $10^{-8}$ (and $10^{-6}$) eV/cell.

To compute the electrons shared and transferred, Bader basins were defined using the Yu-Trinkle algorithm [32] and the corresponding localization (LI($\Omega$)) and delocalization indices ($\delta(\Omega_1, \Omega_2)$) were determined using the DGRID code. [33] The average amount of electron pairs was obtained by integrating the electron pair density $\rho(\Omega_1, \Omega_2)$ over a unique basin ($\Omega_1 = \Omega_2$) whereas $\delta(\Omega_1, \Omega_2)$ were obtained by integrating the non-classical part of the exchange-correlation density (i.e. the same spin electron exchange-correlation density) over distinct basins to provide the number of shared pairs of electrons. The electron population of each atom N($\Omega$) is defined as the sum over the localization and half of the delocalization indices of an atom with all other basins. The same is obtained by simple integration of the electron density over the basin $\Omega$. All compounds analyzed here are binary compounds, i.e. by subtracting the atomic charge to N($\Omega$), we obtain the quantity of electrons transferred between atoms. Further details can be found in the supplementary information of Ref. [7b]. Details concerning the calculation of Z*, $\varepsilon_\infty$ and $\gamma_{TO}$ can be found in the supplementary information of Ref. [7a].

## Supporting Information

Supporting Information is available from the Wiley Online Library or from the author.

## Acknowledgements

We acknowledge the computational resources granted by JARA-HPC from RWTH Aachen University under projects JARA0176, JARA0198 and RWTH0508. The authors gratefully acknowledge support from C. Teichrib and P. Kerres for their support with the thin film preparation and XRR measurements. This work was supported in part by the Deutsche Forschungsgemeinschaft (SFB 917), in part by the Federal Ministry of Education and Research (BMBF, Germany) in the project NEUROTEC (16ES1133 K) and in part by Excellence Initiative of the German federal and state governments (EXS-SF-neuroIC005). J.-Y. R. acknowledges provision from the French Research Funding Agency (ANR-15-CE24-0021-05, SESAME), and computational resources provided by the CÉCI (funded by the F.R.S.-FNRS under Drant No. 2.5020.11) and the Tier-1 supercomputer of the Fédération Wallonie-Bruxelles (infrastructure funded by the Walloon Region under grant agreement n°1117545). S. S. is grateful for a Liebig scholarship from the Verband der Chemischen Industrie e. V., Frankfurt (FCI), and the computational resources provided by the IT center of RWTH Aachen (JARA-HPC project jara0167). S. M. acknowledges furthermore the help and flexible service of the IT Center of the RWTH Aachen University.

## Financial and Conflict of Interest

The authors declare no financial and no conflict of interest.